\documentclass[12pt]{amsart}
\usepackage{amsmath}
\usepackage{latexsym}
\usepackage{amsfonts}
\usepackage{amssymb}
\usepackage{bbm,dsfont}

\newtheorem{proposition}{Proposition}

\newtheorem{lemma}{Lemma}
\newtheorem{corollary}{Corollary}

\theoremstyle{definition}

\newtheorem{definition}{Definition}
\newcommand{\eps}{\varepsilon}
\newcommand{\fii}{\varphi}

\newcommand{\hi}{\mathcal{H}} 
\newcommand{\ki}{\mathcal{K}} 
\newcommand{\lh}{\mathcal{L(H)}} 
\newcommand{\trh}{\mathcal{T(H)}} 
\newcommand{\sh}{\mathcal{S(H)}} 
\newcommand{\eh}{\mathcal{E(H)}} 
\newcommand{\ip}[2]{\left\langle\,#1\,|\,#2\,\right\rangle} 

\newcommand{\kb}[2]{|#1\,\rangle\langle\,#2|} 
\newcommand{\no}[1]{\left\|#1\right\|} 
\newcommand{\tr}{\textrm{tr}} 
\newcommand{\I}{\mathfrak{I}} 

\newcommand{\pfii}{P_{\fii}}
\newcommand{\ppsik}{P_{\psi_k}}
\newcommand{\pphi}{P_{\phi}}
\newcommand{\q}{\mathsf{Q}} 
\newcommand{\p}{\mathsf{P}} 
\newcommand{\qrho}{\mathsf{Q}_\rho} 
\newcommand{\pnu}{\mathsf{P}_\nu} 
\newcommand{\pnud}{\mathsf{P}_{\nu'}} 
\newcommand{\qnu}{\mathsf{Q}_\nu} 
\newcommand{\E}{\mathsf{E}} 
\newcommand{\F}{\mathsf{F}} 
\newcommand{\G}{\mathsf{G}} 
\newcommand{\id}{\mathbbm{1}} 

\newcommand{\f}{\mathcal{F}}

\newcommand{\R}{\mathbb R} 
\newcommand{\RR}{\mathbb R^2}
\newcommand{\C}{\mathbb C} 
\newcommand{\N}{\mathbb N} 
\newcommand{\Z}{\mathbb Z} 
\newcommand{\half}{\frac{1}{2}}

\newcommand{\A}{\mathcal{A}}
\newcommand{\bor}[1]{\mathcal{B}(#1)}
\newcommand{\borel}{\mathcal B(\R)}
\newcommand{\borelr}{\mathcal B(\R^2)}

\newcommand{\ldl}{L^2\left(\mathbb{R},dx\right) } 
\newcommand{\lul}{L^1\left(\mathbb{R}, dx\right) } 
\newcommand{\supp}{{\rm supp}\,}
\newcommand{\diam}{{\rm diam}}
\newcommand{\essup}{{\rm ess}\,{\rm sup}}

\newcommand{\frecc}{\rightarrow}

\begin{document}

\title[Intrinsic unsharpness and approximate repeatability]{Intrinsic unsharpness and approximate repeatability of quantum measurements}

\author[Carmeli]{Claudio Carmeli}
\address{Claudio Carmeli, Dipartimento di Fisica, Universit\`a di
 Genova, Via Dodecaneso 33, 16146 Genova, Italy.} 
\email{carmeli@ge.infn.it}

\author[Heinonen]{Teiko Heinonen}
\address{Teiko Heinonen, Department of Physics, University of Turku, Finland.}
\email{teiko.heinonen@utu.fi}

\author[Toigo]{Alessandro Toigo}
\address{Alessandro Toigo, Dipartimento di Fisica, Universit\`a di
  Genova, and I.N.F.N., Sezione di Genova, Via 
Dodecaneso 33, 16146 Genova, Italy.}
\email{toigo@ge.infn.it}

\maketitle

\section{Introduction}\label{Introduction}

In quantum mechanics, unsharpness has a fundamental role and it has to be taken into account also in theoretical studies. For instance, there is no joint measurement for sharp position and momentum observables. Only unsharp position and momentum observables may allow a joint measurement. Also, every measurement has some effect on the system and hence, an unavoidable disturbance to the subsequent measurements.

In this paper we discuss a quantification of the intrinsic unsharpness of non-discrete observables, such as position and momentum. For this purpose, we introduce the concept of resolution width. It is the minimal size of intervals for which the corresponding effects have suitable low degree of unsharpness. 

It is a well known fact that only discrete observables admit repeatable measurements \cite{Ozawa85}. Hence, non-discrete observables can at best have approximately repeatable measurements \cite{BuLa90b},\cite{DaLe70}. We show that the resolution width is closely connected with the possibility of making approximately repeatable measurements. 

We examine the intrinsic unsharpness and approximate repeatability of position and momentum measurements in detail. We also give a sufficient criterion assuring that discretized versions of position and momentum observables admit repeatable measurements. A necessary inaccuracy relation for any jointly measurable pair of position and momentum observables is formulated using their resolution widths. Joint measurements are closely related to sequential measurements in the sense that a suitable kind of sequential measurement leads to a joint observable. We show that any covariant phase space observable can be formed from a mixture of certain kind of sequential measurements.

We shall proceed as follows. In Section \ref{Accuracy} we give some basic definitions and mathematical facts related to the unsharpness of a quantum observable. The discussion of Subsection \ref{Actual} follows \cite{Busch86} and in Subsection \ref{Resolution} we introduce the notion of resolution width, which is central for everything that follows. In Section \ref{Approximate} we review the definitions and some results on approximately repeatable instruments. Also, the connection between resolution width and approximate repeatability is demonstrated. Sections \ref{Position} and \ref{ApproximatePosition} deal with position measurements. In this concrete case a rather complete analysis can be done. Finally, in Section \ref{Joint} we analyze the role of resolution width and approximate repeatability in joint measurements of position and momentum. 

Concluding this section we fix the notation and recall some basic definitions; for further details we refer to \cite{QTM}, \cite{QTOS},  \cite{PSAQT}. Let $\hi$ be a
complex separable Hilbert space. We denote by $\lh$ and $\trh$ the algebra of bounded
operators and the ideal of trace class operators on $\hi$, respectively. A positive operator $T\in\trh$ of trace one is called a \emph{state} and the set of all states is denoted by $\sh$. A \emph{pure state} is a one-dimensional projection and $\pfii$ denotes the pure state generated by a nonzero vector $\fii\in\hi$. A positive operator bounded from above by the unit operator $\id$ is called an \emph{effect} and the set of all effects is denoted by $\eh$. 

Let $\Omega$ be a nonempty set and $\A$ a $\sigma$-algebra of subsets of $\Omega$. A mapping $\E:\A\to\eh$ is an \emph{observable} if it is $\sigma$-additive with respect to the weak operator topology and $\E(\Omega)=\id$. An observable $\E$ which has only projections in its range, that is, $\E(X)=\E(X)^2$ for any $X\in\A$, is conventionally called a \emph{sharp observable}. We will mostly deal with observables defined on $\borel$, the Borel $\sigma$-algebra of the real line $\R$. 

An \emph{operation} (or \emph{state transformation}) is a positive linear mapping $\Phi:\trh\to\trh$ which satisfies the condition $0\leq \tr[\Phi(T)]\leq 1$ for every $T\in\sh$. An \emph{instrument} is a mapping $X\mapsto\I_X$ from $\borel$ to the set of operations, which satisfies the normalization condition $\tr[\I_{\R}(T)]=1$ for every $T\in\sh$ and is $\sigma$-additive in the sense that, whenever $T\in\trh$ and $(X_i)\subset\borel$ is a sequence of disjoint Borel sets, then
$\I_{\cup_i X_i}(T)=\sum_i \I_{X_i}(T)$ where the sum converges in the trace-norm topology. In order to have a meaningful physical interpretation, it is essential that an operation is completely positive \cite{SEO}. We say that an instrument $\I$ is completely positive if every operation $\I_X, X\in\borel$, is completely positive. This also assures that the instrument is induced by a (normal) premeasurement,  \cite{BuLa90a},\cite{Ozawa84}.

Each instrument $\I$ determines an associated observable $\E$ by the formula
\begin{equation}\label{Ecomp}
\tr[T\E(X)]=\tr[\I_X(T)],\quad X\in\borel,T\in\sh.
\end{equation}
Any instrument satisfying condition (\ref{Ecomp}) is called $\E$-\emph{compatible}.

\section{Intrinsic unsharpness of an observable}\label{Accuracy}

\subsection{Actualizability of effects}\label{Actual}

Let $\E:\A\to\eh$ be an observable and $X\in\A$.
\begin{definition}
An effect $\E(X)$ is \emph{actual} in a state $T$ if
\begin{equation}\label{TE1}
\tr[T\E(X)]=1.
\end{equation}
An effect which is actual in some state is \emph{actualizable}. 
\end{definition}

Condition (\ref{TE1}) means that a measurement outcome belongs to the set $X$ with probability 1 when a measurement of the observable $\E$ is performed in the state $T$.

An actualizable effect is actual in some pure state. Indeed, assume that an effect $\E(X)$ is actual in a mixed state $T$. The state $T$ has a (trace norm convergent) $\sigma$-convex decomposition of the form
\begin{equation}\label{decom1}
T=\sum_{i=1}^{\infty} p_i T_i,
\end{equation}
where $(p_i)$ is a sequence of positive numbers summing to 1 and $(T_i)$ is a sequence of pure states. We then have
\begin{equation*}
1=\tr[T\E(X)]=\sum_{i=1}^{\infty} p_i \tr[T_i\E(X)],
\end{equation*}
which implies that $\tr[T_i\E(X)]=1$ for every $i=1,2,\ldots$. For the reader's convenience we give a proof for the following elementary fact. 

\begin{proposition}\label{prop:eigen1}
An effect $\E(X)$ is actualizable if and only if it has eigenvalue 1.
\end{proposition}

\begin{proof}
If $\E(X)$ has eigenvalue 1 and $\fii$ is a corresponding eigenvector, then $\E(X)$ is actual in the state $\pfii$. 

Now, assume that $\E(X)$ is an actualizable effect. Then there is a pure state $T$ such that (\ref{TE1}) holds. This also means that there is a unit vector $\psi\in\hi$ such that 
\begin{equation}\label{psiE1}
\ip{\psi}{\E(X)\psi}=1.
\end{equation}
Using the Cauchy-Schwarz inequality and the fact that $\E(X)\leq\id$ we get
\begin{equation*}
1=|\ip{\psi}{\E(X)\psi}|\leq \no{\psi}\no{\E(X)\psi}=\no{\E(X)\psi}\leq 1,
\end{equation*}
and therefore,
\begin{equation*}
|\ip{\psi}{\E(X)\psi}| = \no{\psi}\no{\E(X)\psi}.
\end{equation*}
This implies that the vector $\E(X)\psi$ is a scalar multiple of $\psi$, i.e., $\E(X)\psi=\alpha \psi$ for some $\alpha\in\C$. It then follows from (\ref{psiE1}) that $\alpha=1$.
\end{proof}

Every nonzero projection is an actualizable effect. It is easy to construct also other examples. For instance, take two orthogonal unit vectors $\fii_1$ and $\fii_2$ and fix $0<p<1$. Then the effect $P_{\fii_1}+pP_{\fii_2}$ is actualizable but not projection. Generally, however, actualizability is a strong requirement and often not fulfilled. Therefore, the following weakening is needed.  

\begin{definition}
Let $\frac{1}{2} \leq c < 1$.  An effect $\E(X)$ is $c$-\emph{actual} in a state $T$ if
\begin{equation}\label{TEc}
\tr[T\E(X)] > c.
\end{equation}  
An effect which is $c$-actual in some state is $c$-\emph{actualizable}. 
\end{definition}

The reason for the restriction $c \geq \frac{1}{2}$ is to avoid the situation where an effect $\E(X)$ and its complement $\E(\R\smallsetminus X)=\id-\E(X)$ would both be $c$-actual in the same state. Moreover, since $c$-actualizability is introduced as an approximation of actualizability, $c$ can usually be thought as a number close to 1. 

Since $\E(X)$ is a positive operator, the operator norm can be expressed as 
$$
\no{\E(X)}=\sup_{T\in\sh} \tr[T\E(X)].
$$
This leads to the following conclusion.

\begin{proposition}\label{prop:normc}
An effect $\E(X)$ is $c$-actualizable if and only if $\no{\E(X)}>c$.
\end{proposition}  

Assume that an effect $\E(X)$ is $c$-actual in a state $T$. Using a $\sigma$-convex decomposition as in (\ref{decom1}) for $T$, it is seen that there is a pure state $T_i$ such that $\E(X)$ is $c$-actual in $T_i$. However, unlike the case of an actualizable effect, there may be a pure state $T_i$ in the decomposition of $T$ such that $\E(X)$ is not $c$-actual in $T_i$.

\begin{definition}
An effect $\E(X)$ is \emph{almost actualizable} if it is $c$-actualizable for every $\frac{1}{2}\leq c<1$.
\end{definition}

As a direct consequence of Proposition \ref{prop:normc} we conclude that an effect $\E(X)$ is almost actualizable if and only if $\no{\E(X)}=1$. If an effect $\E(X)$ is almost actualizable but not actualizable, then 1 belongs to the spectrum of $\E(X)$ but it is not an eigenvalue.

The difference between actualizability and almost actualizability has been pointed out, for instance, by Ballentine in \cite[footnote 4]{Ballentine70}. It seems quite impossible to distinguish between actualizability and almost actualizability in any practical situation. Therefore, we take almost actualizability to represent the optimal reality content which an effect can have.

However, it is an interesting fact that in some cases the theoretical difference between actualizability and almost actualizability is crucial. A physically relevant example is the canonical phase observable, whose all nontrivial effects are almost actualizable but not actualizable; see \cite{BuLaPeYl01} and \cite{HeLaPePuYl03}. Other interesting examples are the localization observables of a massless particle with non-zero helicity constructed by Castrigiano in \cite{Castrigiano81}. He showed that for these observables any effect corresponding to a bounded Borel set with non-void interior is almost actualizable but not actualizable.

\subsection{Resolution width}\label{Resolution}

In the rest of the paper any observable $\E$ in consideration is, if not otherwise stated,  defined on $\borel$. In later sections we study position and momentum observables, which have the same null sets as the Lebesgue measure. For our purposes in this section, it is enough to assume that each observable $\E$ has the whole real line $\R$ as its support. This assumption is equivalent to the condition that $\E(I)\neq O$ for every open interval $I\subset\R$. With some simple modifications one could make a similar analysis for observables which are supported in an interval.

Let $X\in\borel$ and assume that $\E(X)$ is a $c$-actualizable effect for some fixed $c$. If $Y\in\borel$ is such that $X\subseteq Y$, then $\E(X)\leq \E(Y)$ and therefore, also the effect $\E(Y)$ is $c$-actualizable. With this in mind, we may ask for the minimal width such that any effect $\E(I)$ corresponding to an interval $I$ bigger than this width is $c$-actualizable.

For any $x\in\R, r\in\R_+$, we denote the open interval
$(x-\frac{r}{2},x+\frac{r}{2})$ by $I_{x;r}$. 

\begin{definition}\label{gammaEc}
Let $\frac{1}{2} \leq c < 1$. We denote
\begin{equation*}
\gamma(\E;c):=\inf \{ r>0\mid \E(I_{x;r}) \textrm{ is $c$-actualizable for every $x\in\R$}\},
\end{equation*}
and say that $\gamma(\E;c)$ is the \emph{resolution width of} $\E$ \emph{with confidence level} $c$.
\end{definition}

We adopt the definition $\inf\emptyset =\infty$, and thus, the range of possible values of $\gamma(\E;c)$ is the closed interval $[0,\infty]$. The function $c\mapsto \gamma(\E;c)$ from $[ \half, 1 )$ to $[ 0,\infty ]$ is increasing, that is, 
\begin{equation}\label{monotone}
c_1\leq c_2\ \Rightarrow \gamma(\E;c_1)\leq\gamma(\E;c_2).
\end{equation}
It is natural to give the following definition.
\begin{definition}\label{gammaE1}
We denote
\begin{equation*}
\gamma(\E;1):=\lim_{c\to 1-}\gamma(\E;c),
\end{equation*}
and say that $\gamma(\E;1)$ is the \emph{resolution width of} $\E$ \emph{with confidence level} $1$.
\end{definition}

\begin{proposition}
\begin{eqnarray}
\label{gauno}
\gamma(\E;1)=\inf \{ r>0\mid \E(I_{x;r}) \textrm{ is almost actualizable for every $x\in\R$}\}.
\end{eqnarray}
\end{proposition}

\begin{proof}
Let us first note that by (\ref{monotone}), we can write Definition \ref{gammaE1} alternatively as
$\gamma\left(\E;1\right)  =  \sup_{\frac{1}{2}\leq c < 1} \gamma\left(\E,c\right)$, while the right hand side of equation (\ref{gauno}) can be rewritten as $ \inf \{ r>0\mid \no{\E(I_{x;r})}=1\,\forall \, x\in\R \}=:M$. From this it is evident that $\gamma\left(\E;c\right)\leq M$ for each $\frac{1}{2}\leq c <1$.
 Hence, $\gamma\left(\E;1\right)={\rm sup}_{\frac{1}{2}\leq c <1} \gamma\left(\E;c\right)\leq M$.
Fix now $\delta>0$, then $\gamma(E,1)+\delta> \gamma(E,c)$ for each $\frac{1}{2}\leq c <1$. It follows that $\no{\E(I_{x;\gamma(E,1)+\delta})}> c$ for every  $\frac{1}{2}\leq c <1$ so that $\no{\E(I_{x;\gamma(E,1)+\delta})}=1$. This being true for each $\delta>0$ we can  conclude that 
  $\gamma\left(\E,1\right)\geq M$, and the claim now follows.
\end{proof}

The function $\gamma(\E;\cdot)$ is a description of the intrinsic unsharpness, or inaccuracy, of the observable $\E$. Typically, a single number $\gamma(\E;c)$ with a well chosen confidence level $c$ (or a finite sample) gives enough information on the precision of $\E$.

The best resolution width $\gamma(\E;1)=0$ is achieved, for instance, when $\E$ is a sharp observable. Also the worst case is possible, namely, that $\gamma(\E;\half)=\infty$. To give an example of this latter situation, let $\lambda$ be a probability measure on $\borel$ and define an observable $\E$ by formula $\E(X)=\lambda(X)\id$. If $I$ and $J$ are two disjoint intervals, then either $\lambda(I)\leq\half$ or $\lambda(J)\leq\half$. This implies that $\gamma(\E;\half)=\infty$.

\section{Approximately repeatable instruments}\label{Approximate}

A measurement is said to be repeatable if its repetition does not give a new result (from a probabilistic point of view). The quantum theory of sequential measurements leads naturally to the following formulation of repeatability; see, for instance, \cite{BuCaLa90}.

\begin{definition}\label{repeatable}
An instrument $\I$ is \emph{repeatable} if for all
$T\in\sh$ and $X\in\borel$,
\begin{equation*}
\tr \left[ \I_{X}\left( \I_X(T) \right)
  \right] = \tr[\I_X(T)].
\end{equation*}
\end{definition}

It is a well known result that an instrument $\I$ can be repeatable only if its associated observable $\E$ is discrete \cite{Ozawa85}, that is, there is a countable subset $X\subset\R$ such that $\E(X)=\id$. Under this precondition, a necessary and sufficient requirement in order that there exists an $\E$-compatible repeatable instrument is that all the nonzero effects $\E(X)$ are actualizable; see, for instance, \cite[Section II.3.5]{OQP}.

To understand the properties and operational meaning of non-discrete observables, one is forced to seek alternatives to Definition \ref{repeatable}. To formulate two existing proposals, we denote for each $X\subseteq\R$ and $\eps > 0$,
\begin{equation*}
X_{\eps}:=\bigcup_{x\in X} I_{x;\eps}=\{y\in\R\mid |x-y|<\frac{\eps}{2} \textrm{ for some } x\in X \}.
\end{equation*}

\begin{definition}
Let $\I$ be an instrument, $\eps > 0$ and $\half \leq c < 1$.
\begin{itemize}
\item[(i)] $\I$ is $\eps$-\emph{repeatable} if for all
$T\in\sh$ and $X\in\borel$,
\begin{eqnarray}
\tr \left[ \I_{X_{\eps}}\left( \I_X(T) \right)
  \right] & = & \tr[\I_X(T)].
\end{eqnarray}
\item[(ii)] $\I$ is $(\eps,c)$-\emph{repeatable} if for all
$T\in\sh$ and $X\in\borel$ such that $\tr[\I_X(T)]\neq 0$,
\begin{eqnarray}\label{epsc}
\tr \left[ \I_{X_{\eps}}\left( \I_X(T) \right)
  \right] & > & c \cdot \tr[\I_X(T)].
\end{eqnarray}
\end{itemize}
\end{definition}

Here we clearly have a chain of properties: repeatability implies $\eps$-repeatability, which, in turn, implies $(\eps,c)$-repeatability.

The concept of $\eps$-repeatable instrument was introduced by Davies and Lewis in \cite{DaLe70} to replace the repeatability condition for non-discrete observables. They proved that if $\E$ is an observable such that any effect $\E(I)$ corresponding to an interval $I$ is actualizable, then for each $\eps >0$, there exists an $\E$-compatible instrument which is $\eps$-repeatable \cite[Theorem 4]{DaLe70}.

As $\eps$-repeatability requires that the associated observable has actualizable effects, one needs more relaxed concept for general investigations. The important definition of an $(\eps,c)$-repeatable instrument was introduced in \cite{BuLa90b} and \cite{BuCaLa90}; see also \cite[Section IV.1]{OQP} and \cite{BuGrLa95b}. 

Assume that an observable $\E$ admits an $(\eps,c)$-repeatable instrument $\I$. This implies that whenever $\E(X)\neq O$, the effect $\E(X_\eps)$ is $c$-actualizable. Indeed, choose a state such that $\tr[T\E(X)]\neq 0$. Then condition (\ref{epsc}) implies that $\tr[T_X\E(X_{\eps})]>c$, where $T_X:=\I_X(T)/\tr[\I_X(T)]$. 

\begin{proposition}\label{prop:noepsc}
Let $\E$ be an observable whose support is $\R$. If $\eps < \gamma(\E;c)$, there is no $\E$-compatible instrument which is $(\eps,c)$-repeatable.
\end{proposition}

\begin{proof}
Fix $\eps'$ such that $\eps <\eps' < \gamma(\E;c)$. By Definition \ref{gammaEc}, there is $x\in\R$ such that $\tr[T\E(I_{x;\eps'})]\leq c$ for every $T\in\sh$. Choose $X=I_{x;\eps'-\eps}$, in which case $X_{\eps}=I_{x;\eps'}$. Then $\E(X)\neq O$ but $\E(X_{\eps})$ is not $c$-actualizable. This means, due to the discussion in the previous paragraph, that there cannot be any $(\eps,c)$-repeatable instrument.
\end{proof}

The following positive result on the existence of $(\eps,c)$-repeatable instruments is a modification of Theorem 4 in \cite{DaLe70}.

\begin{proposition}\label{prop:epsc}
Let $\E$ be an observable, $\half\leq c < 1$ and $\eps > 2\cdot \gamma(\E;c)$. Then there is a completely positive $\E$-compatible instrument which is $(\eps,c)$-repeatable.
\end{proposition}  

\begin{proof}
For each $n\in\Z$, denote by $X_n$ the half-open interval $[\frac{n}{2}\eps,\frac{(n+1)}{2}\eps)$ and choose a pure state $P_{\psi_n}$ such that the effect $\E(X_n)$ is $c$-actual in the state $P_{\psi_n}$. The formula 
\begin{equation*}
\I_X(T):=\sum_{n=-\infty}^{\infty} \tr[T\E(X\cap X_n)]\ P_{\psi_n}
\end{equation*}
defines an $\E$-compatible instrument $\I$. 

Fix an orthonormal basis $\{\varphi_k\}$ for $\hi$. Expanding the trace in this basis, each $\I_X, X\in\borel$, can be written in the Kraus form
\begin{equation*}
\I_X(T)=\sum_{k,n=-\infty}^{\infty} A_{k,n} T A_{k,n}^\ast,
\end{equation*}
where 
\begin{equation*}
A_{k,n}:=\kb{\psi_n}{\E(X\cap X_n)^\half \varphi_k}.
\end{equation*}
Thus, the instrument $\I$ is completely positive; see, e.g., \cite[\textsection 3, Theorem 1]{SEO}.

To prove that $\I$ is $(\eps,c)$-repeatable, let $X\in\borel$ and $T\in\trh$. We then get
\begin{eqnarray*}
\tr \left[ \I_{X_{\eps}}\left( \I_X(T) \right) \right] &=&
\sum_{n=-\infty}^{\infty} \sum_{k=-\infty}^{\infty} \tr[\ppsik\E(X_{\eps}\cap X_n)]\ \tr[T\E(X\cap X_k)] \\
&=& \sum_{k=-\infty}^{\infty} \tr[\ppsik\E(X_{\eps})]\ \tr[T\E(X\cap X_k)]
\end{eqnarray*}
If $X\cap X_k\neq\emptyset$, then $X_k\subseteq X_{\eps}$. This implies that either $\tr[T\E(X\cap X_k)]=0$ or $\tr[\ppsik\E(X_{\eps})] > c$. Assume now that $\tr[T\E(X)]\neq 0$, in which case $\tr[T\E(X\cap X_k)]\neq 0$ at least for some $k$. Therefore, 
\begin{eqnarray*}
\sum_{k=-\infty}^{\infty} \tr[\ppsik\E(X_{\eps})]\ \tr[T\E(X\cap X_k)] & > &\sum_{k=-\infty}^{\infty} c \cdot\tr[T\E(X\cap X_k)] \\
&=& c \cdot \tr[\I_X(T)].
\end{eqnarray*}

\end{proof}

\section{Intrinsic unsharpness of position observables}\label{Position}

\subsection{Definition of position observables}

In the rest of this paper $\hi=\ldl$. The canonical position observable, denoted by $\q$, is the sharp observable defined as
\begin{equation*}
[\q(X)\psi](x)=\chi_X(x)\psi(x),\quad X\in\borel,
\end{equation*}
where $\chi_X$ is the characteristic function of $X$.

Let $\rho$ be a probability measure on $\R$. The formula
\begin{equation}\label{qrho}
\qrho(X)=\int \rho(X-x)\ d\q(x),\quad X\in\borel,
\end{equation}
defines an observable $\qrho$, whose action on a function $\psi\in\hi$ is given by
\begin{equation}\label{multiplica}
[\qrho(X)\psi](x)=\rho(X-x)\psi(x).
\end{equation}
We call $\qrho$ a \emph{position observable}; motivation for this terminology is briefly explained below. Note that the canonical position observable $\q$ is recovered from equation (\ref{qrho}) when $\rho=\delta_0$, the Dirac measure concentrated at the origin.

The observable $\qrho$ has the same kinematical symmetry properties as the canonical position observable $\q$. Namely, for every $q,p\in\R$, define the unitary operators $U_q$ and $V_p$ by
\begin{eqnarray*}
\left( U_q \psi \right) (x) &=& \psi(x-q), \\
\left( V_p \psi \right) (x) &=& e^{ipx} \psi(x).
\end{eqnarray*}
These unitary operators correspond to position shift and momentum boost, respectively.
The kinematical symmetry properties of $\qrho$ can be expressed as
\begin{eqnarray}
U_q \qrho(X) U_q^* &=& \qrho(X+q), \label{poscov}\\
V_p \qrho(X) V_p^* &=& \qrho(X). \label{posinv}
\end{eqnarray}
As proved in \cite[Proposition 1]{CaHeTo04} and \cite[Proposition 3]{CaHeTo05}, the observables satisfying the symmetry conditions (\ref{poscov}) and (\ref{posinv}) are in one-to-one correspondence with the probability measures on $\R$ via the formula (\ref{qrho}). 

An observable $\qrho$ can be interpreted as an imprecise version or a smearing of the canonical position observable $\q$, the probability measure $\rho$ quantifying the inaccuracy. 
We refer to \cite{OQP} and \cite{CRQM} for discussions on the interpretation and properties of $\qrho$.

\subsection{Resolution width of a position observable}\label{ResolutionPosition}

Since a position observable has the simple form (\ref{qrho}), we can express the corresponding resolution width $\gamma(\qrho;c)$ in terms of the probability measure $\rho$. We denote by $\essup_{x\in\R} f(x)$ the essential supremum of a function $f:\R\to\R$ with respect to Lebesgue measure.

\begin{proposition}\label{prop:gammaEQc}
Let $\qrho$ be a position observable and $\half\leq c <1$. Then 
\begin{equation}\label{gammaEQc}
\gamma(\qrho;c)=\inf\{r > 0 \mid \essup_{x\in\R}\rho(I_{x;r}) > c\}
\end{equation}
and this is a finite number. 
\end{proposition}

\begin{proof}
The covariance condition (\ref{poscov}) implies that for any $x\in\R$,
\begin{equation}\label{estimation}
\no{\qrho(I_{x;r})}=\no{U_x\qrho(I_{0;r})U_x^*}=\no{\qrho(I_{0;r})}.
\end{equation}
Since $\qrho(I_{0;r})$ is a multiplicative operator in $\ldl$, we have 
\begin{equation*}
\no{\qrho(I_{0;r})}=\essup_{x\in\R}\rho(I_{0;r}+x)=\essup_{x\in\R}\rho(I_{x;r}),
\end{equation*}
and hence, the resolution width $\gamma(\qrho;c)$ has the claimed form.

As
\begin{equation*}
\lim_{r\to\infty} \rho(I_{0;r})=\rho(\R)=1,
\end{equation*}
there is an interval $I_{0;R}$ such that $\rho(I_{0;R})>c$. Fix $\delta>0$. Then for every $x\in I_{0;\delta}$, we have $I_{0;R}\subseteq I_{x;R+\delta}$ and hence, $\rho(I_{x;R+\delta})>c$. Therefore,
\begin{equation*}
\essup_{x\in\R} \rho(I_{x;R+\delta}) \geq \essup_{x\in I_{0;\delta}} \rho(I_{x;R+\delta}) >c.
\end{equation*}
This shows that $\gamma(\qrho;c)\leq R$.
\end{proof}

Let us note that the translation covariance of $\qrho$ makes an estimation of the resolution width more achievable than for observables in general. Indeed, if one finds a state $T$ such that the effect $\qrho(I_{0;r})$ is $c$-actual in $T$, then by equation (\ref{estimation}) one concludes that $\gamma(\E;c)\leq r$.

To formulate the following result, we denote by $\diam X$ the diameter of a set $X\subseteq\R$, i.e., $\diam X:=\sup \{ |x-y| \mid x,y\in X \}$. We also recall that the support of a probability measure $\lambda$ on $\borel$ can be expressed as $\supp\lambda=\cap \{ X\subseteq\R \mid \textrm{$X$ closed, $\lambda(X)=1$} \}$.

\begin{proposition}\label{prop:gammaEQ1}
Let $\qrho$ be a position observable. Then 
\begin{equation}\label{gammaEQ1}
\gamma(\qrho;1) = \diam\,\supp \rho.
\end{equation}
\end{proposition}

The proof of Proposition \ref{prop:gammaEQ1} follows easily from the next lemma. We emphasize that $\gamma(\qrho;1)$ may be infinite.

\begin{lemma}
Let $r>0$.
\begin{itemize}
\item[(i)] If $\essup_{x\in \R} \rho(I_{x;r}) = 1$, then ${\rm diam}\,\supp \rho \leq r$.
\item[(ii)] If ${\rm diam}\,\supp \rho < r$, then $\essup_{x\in \R} \rho(I_{x;r}) = 1$.
\end{itemize}
\end{lemma}
\begin{proof}
\begin{itemize}
\item[(i)] Suppose $\essup_{x\in \R} \rho(I_{x;r}) = 1$.
For each $\alpha > 0$, denote
\begin{equation*}
B_\alpha = \left\{ x\in\R \mid \rho (I_{x,r}) \geq 1 - \alpha \right\}.
\end{equation*}
Since $\rho$ is bounded, $B_\alpha $ is a bounded set.
We then have $B_\alpha \neq \emptyset$ and $B_\alpha \subset B_\beta$ if $\alpha < \beta$.  Choose $x_n \in B_{1/n}$ $\forall n \in \N$. Let $\left( x_{n_k} \right)_{n\in \N}$ be a convergent subsequence, and let $\overline{x}$ be its limit. Thus, $\forall \delta > 0$ $\exists k_\delta \in \N$ such that $k \geq k_\delta$ implies $I_{\overline{x}, r+\delta} \supset I_{x_{n_k},r}$. So, $\rho \left( I_{\overline{x}, r+\delta} \right) \geq \rho \left( I_{x_{n_k},r} \right) \geq 1 - 1/n_k$ for all $k \geq k_\delta$. It follows that
$\rho \left( I_{\overline{x}, r+\delta} \right) = 1$. Hence
$$
\rho \left( \overline{I_{\overline{x}, r}} \right) =
\rho \left( \cap_{\delta > 0} I_{\overline{x}, r+\delta} \right) = \lim_{\delta \to 0^+}
\rho \left( I_{\overline{x}, r+\delta} \right) = 1,
$$
i.e.~$\supp \rho \subset \overline{I_{\overline{x}, r}}$.
\item[(ii)] Suppose ${\rm diam}\,\supp \rho = r^{\prime} < r$. Let $\overline{x}$ be such that $\supp \rho \subset \overline{I_{\overline{x}, r^{\prime}}}$. Since $\overline{I_{\overline{x}, r^{\prime}}} \subset I_{x, r}$ for all $x\in \left(\overline{x} - (r-r^{\prime})/2, \overline{x} + (r-r^{\prime})/2 \right)$, so that $\rho (I_{x,r}) = \rho (\overline{I_{\overline{x}, r^{\prime}}}) = 1$ for such $x$'s, and the interval $\left(\overline{x} - (r-r^{\prime})/2, \overline{x} + (r-r^{\prime})/2 \right)$ has nonzero Lebesgue measure, we have $\essup_{x\in\R} \rho (I_{x,r}) = 1$.
\end{itemize}
\end{proof}

\begin{proof}[Proof of Proposition~\ref{prop:gammaEQ1}]
Let $r = \gamma(\qrho;1)$ (possibly $r=\infty$). Then $r$ is fixed by the conditions: (a) $\essup_{x\in\R} \rho (I_{x,r^{\prime}}) = 1$ for all $r^{\prime} > r$; (b) $\essup_{x\in\R} \rho (I_{x,r^{\prime}}) < 1$ for all $r^{\prime} < r$ (if $r=\infty$, condition (a) is trivial).
From (a) and item (i) in the lemma, ${\rm diam}\,\supp \rho \leq r$ follows. Now suppose ${\rm diam}\,\supp \rho = r^{\prime} < r$. Let $\eps > 0$ be such that $r^{\prime} < r^{\prime} + \eps < r$. By item (ii) in the lemma, $\essup_{x\in\R} \rho (I_{x,r^{\prime} + \eps}) = 1$, a contradiction. Hence, ${\rm diam}\,\supp \rho = r$.
\end{proof}

It is a direct consequence of Proposition \ref{prop:gammaEQ1} that the best resolution width $\gamma(\qrho; 1)=0$ is achieved only if $\rho$ is the Dirac measure $\delta _{\overline{x}}$ for some $\overline{x}\in\R$, which is the case exactly when $\qrho$ is a sharp position observable; see also \cite[Proposition 2]{CaHeTo04} for a related characterization. A natural relaxation is to require that $\gamma(\qrho; 1)$ is a finite (but nonzero) number. In this case the uncertainty of a measurement result can be made negligible whenever outcome sets are bigger than $\gamma(\qrho; 1)$. Another interesting possibility is that $\gamma(\qrho; c)=0$ for some $\half\leq c<1$. This means that a measurement of $\qrho$ is efficient enough to discriminate arbitrarily small intervals if uncertainty of $1-c$ is tolerated. This situation is  characterized in Propositions \ref{prop:EQc0}.

\begin{proposition}\label{prop:EQc0}
Let $\qrho$ be a position observable and $\half \leq c < 1$. The following conditions are equivalent: 
\begin{itemize}
\item[(i)] $\gamma(\qrho; c)=0$;
\item[(ii)]  there exists $\overline{x}\in \R$ and a probability measure
 $\lambda$ with 
 $\overline{x}\in {\rm supp\,}\lambda $ such that 
 \begin{equation}\label{convex}
 \rho =c \delta _{\overline{x}}+(1-c)\lambda.
 \end{equation}
\end{itemize}
\end{proposition}

\begin{proof}
Assume that (i) holds. By Proposition \ref{prop:gammaEQc} this means that
\begin{equation}\label{reg1}
\forall r > 0:\ \essup_{x\in\R} \rho (I_{x;r})>c.
\end{equation}
For each $r >0$, denote
$$
A_{r}=\{x\in\R \mid \rho(I_{x;r})>c\}.
$$
Since $r_1< r_2$ implies $A_{r_1}\subseteq
A_{r_2}$, it follows from (\ref{reg1}) that $A_r\neq\emptyset$ for every $r>0$. For each $n=1,2\dots$, we choose an
element $x_n\in A_{1/n}$. We then have $\rho(I_{x_n;1/n})>c$, and
since $\rho$ is a finite measure, the sequence $(x_n)_{n\geq 1}$ is
bounded. Hence, there exists a subsequence
$(x_{n_k})_{k\geq 1}$ converging to some $\overline{x}\in \R$. 
For each $\beta >0$, there exists $k$ such
that $I_{x_{n_k};1/{n_k}}\subset I_{\overline{x};\beta}$, so that $\rho(I_{\overline{x};\beta})>c$.
Thus,
$$
\rho(\{ \overline{x} \}) = \rho(\cap _{\beta>0}
I_{\overline{x};\beta}) = \lim_{\beta \to 0}
\rho(I_{\overline{x};\beta}) \geq c. 
$$
It follows that $\lambda = (1-c)^{-1}\rho - c (1-c)^{-1} \delta_{\overline{x}}$ is a
probability measure. For any $\beta >0$, we have $\lambda
(I_{\overline{x};\beta}) = (1-c)^{-1} \left( \rho(I_{\overline{x};\beta}) - c \right) >0$,
which implies that $\overline{x}\in {\rm supp}\lambda$. Thus, (i) implies (ii).

Assume that (ii) holds and let $r > 0$. Then
\begin{eqnarray*}
\essup_{x\in\R} \rho (I_{x;r})
& = & \essup_{x\in\R} \left\{ c \chi_{I_{\overline{x},r}} (x) + (1-c)\lambda\left(I_{x,r}\right)  \right\} \\
& \geq & \essup_{x\in I_{\overline{x},r}} \left\{ c \chi_{I_{\overline{x},r}} (x) + (1-c)\lambda\left(I_{x,r}\right)  \right\} \\
& = & c + (1-c)\cdot \essup_{x\in I_{\overline{x},r}} \lambda\left(I_{x,r}\right).
\end{eqnarray*}
Since $\overline{x} \in {\rm supp\,} \lambda$, it follows that $\lambda(I_{\overline{x},r /2}) = \eps >0$. For any $x\in I_{\overline{x},r /2}$, we have $I_{x,r} \supset I_{\overline{x},r /2}$, and therefore $\lambda(I_{x,r}) \geq \eps$. Thus,
$$
\essup_{x\in\R} \rho (I_{x;r}) \geq c + (1-c) \eps > c,
$$
which means that $\gamma(\qrho;c) \leq r$. As this holds for every $r >0$, we get (i).
\end{proof}

Proposition \ref{prop:EQc0} shows, especially, that $\gamma(\qrho; c)=0$ can hold only if $\qrho$ is a mixture (convex combination) of a sharp position observable and some other position observable.

\section{Approximate repeatability of position measurements}\label{ApproximatePosition}

Since we have shown in Proposition \ref{prop:gammaEQc} that $\gamma(\qrho;c)$ is a finite number for every $\half\leq c<1$, it follows from Proposition \ref{prop:epsc} that $\qrho$ admits $(\eps,c)$-repeatable instruments for any confidence level $\half\leq c<1$ whenever $\eps$ is chosen big enough.  In Subsection \ref{Covariant} we show that a position observable admits also a covariant instrument which has better approximate repeatability property than the one used in the proof of Proposition \ref{prop:epsc}. In Subsection \ref{Discrete} we discuss the possibility of discretizing a position observable to achieve repeatability.  

\subsection{Approximately repeatable covariant instrument}\label{Covariant}

Let $\E$ be an observable, $T_0\in \sh$, and define
\begin{equation}\label{IT0}
\I_X (T) := \int_X U_x T_0 U_x^\ast \ \tr[T\E(dx)],\qquad X\in\borel, T\in\sh.
\end{equation}
This formula defines an $\E$-compatible instrument $\I$. 

\begin{proposition}
The instrument $\I$ is completely positive.
\end{proposition}

\begin{proof}
Let $X\in\borel$. The dual mapping $\I_X^\ast$ of $\I_X$ is
$$
\I_X^\ast (B) = \int_X \tr [U_x T_0 U_x^\ast B] d\E(x), \qquad B\in\lh.
$$
The mapping $\I_X$ is completely positive exactly when $\I_X^\ast$ is completely positive, and thus, we need to show that $\I_X^\ast$ is $N$-positive for each $N=1,2,\ldots$; see, e.g., \cite[\textsection 2]{SEO}. Fix $N$ and let $\psi _i\in\hi$ and $B_{jk}\in \lh$ for $1\leq i,j,k\leq N$. Then 
\begin{eqnarray}
&&\sum\nolimits_{ijk}\ip{\psi _{i}}{\I_X^{\ast }\left( B_{ji}^{\ast }B_{jk}\right) \psi _{k}} \nonumber \\
&&\qquad =\sum\nolimits_{ik}\int_X
\tr\left[ \sum\nolimits_{j}B_{jk}U_x
T_{0} U_x^\ast B_{ji}^{\ast }\right] \ip{\psi _{i}}{ d\E (x) \psi _{k}} \label{Mah...}
\end{eqnarray}
By the Naimark dilation theorem, there exist a Hilbert space $\widetilde{\hi}$, an isometry $W: \hi \frecc \widetilde{\hi}$, and a sharp observable $\F:\borel\to\mathcal{E}(\widetilde{\hi})$ such that $\E(X) = W^\ast \F(X) W $ for all $X\in\borel$. It is not restrictive to assume that  $\widetilde{\hi}$ is the Hilbert space $L^2 (\R, \mu; \ki)$,  where $\mu$ is a Borel measure on $\R$, $\ki$ is an infinite dimensional Hilbert space $\ki$, and $\F$ is given by
$$
[\F(X) \phi] (x) = \chi_X (x) \phi(x).
$$
(This follows from the fact that we don't assume the dilation to be minimal. For the relevant form of the spectral theorem, see e.g. \cite[Section IX.10]{CFA}.)
We thus have
$$
\ip{\psi _{i}}{ d\E (x) \psi _{k}} =
\ip{(W \psi _{i}) (x)}{(W \psi _{k}) (x)} d\mu(x),
$$
and the right hand term in the equation (\ref{Mah...}) can be written as
\begin{eqnarray*}
&& \int_X \sum\nolimits_{ik}\tr\left[ \sum\nolimits_{j}B_{jk}U_x T_{0} U_x^\ast B_{ji}^{\ast }\right] \ip{(W \psi _{i}) (x)}{(W \psi _{k}) (x)} d\mu(x) \\
&&\qquad =\int_X 
\tr \left[ C\left( x\right) D\left( x\right) \right] d\mu(x),
\end{eqnarray*}
where for each $x\in\R$ we have introduced the $N\times N$-matrices $C\left( x\right) $
and $D\left( x\right) $
\begin{eqnarray*}
C\left( x\right) _{ki} &=&\tr\left[\sum\nolimits_{j}B_{jk}U_x
T_{0}U_x^\ast B_{ji}^{\ast }\right]  \\
D\left( x\right)_{ik} &=&\ip{(W \psi _{i}) (x)}{(W \psi _{k}) (x)} .
\end{eqnarray*}
Since these matrices are positive semidefinite, we have
\begin{equation*}
\int \tr \left[ C\left( x\right) D\left( x\right) \right] d\mu(x) = \int \tr\left[ C\left( x\right) ^{\half}D\left( x\right) C\left(x\right) ^{\half}\right] d\mu(x)\geq 0,
\end{equation*}
and the claim follows.
\end{proof}

The instrument $\I$ satisfies the covariance condition
\begin{equation*}
U_q\I_X(U_q^\ast TU_q)U_q^\ast=\I_{X+q}(T),\qquad q\in\R,X\in\borel, T\in\sh,
\end{equation*}
exactly when the associated observable $\E$ is translation covariant. Davies has proved that if $\E$ is a translation covariant sharp observable, then any $\E$-compatible covariant instrument has the form (\ref{IT0}) for some $T_0\in\sh$ \cite[Theorem 3]{Davies70}. Moreover, as noted by Busch and Lahti in \cite[Section 3.1]{BuLa90b}, a translation covariant sharp observable has $\eps$-repeatable instrument of the form (\ref{IT0}) for each $\eps>0$.
They also pointed out that any position observable has $(\eps,c)$-repeatable instrument of the type (\ref{IT0}) for suitable numbers $\eps$ and $c$. In the next proposition we make this observation explicit using the concept of resolution width.

\begin{proposition}
Let $\qrho$ be a position observable and $\half\leq c<1$. For each $\eps > \gamma(\qrho;c)$, there is a $\qrho$-compatible instrument of the form (\ref{IT0}) which is $(\eps,c)$-repeatable.
\end{proposition}

\begin{proof}
Since $\eps > \gamma(\qrho;c)$, there is a state $T_0\in \sh$ such that $\tr \left[ T_0 \qrho (I_{0,\eps}) \right] > c$.
Let $\I$ be the instrument generated by $T_0$. For any $T\in\sh$ and $X\in\borel$, we get
\begin{eqnarray*}
\tr\left[ \I_{X_\eps} \left(\I_X (T) \right)\right] &=& \tr \left[ \I_X (T) \qrho (X_\eps) \right] \\
&=& \int_X \tr \left[ U_x T_0 U_x^\ast \qrho (X_\eps) \right] \ \tr[T\qrho(dx)] \\
&=& \int_X \tr \left[ T_0 \qrho (X_\eps - x) \right] \ \tr[T\qrho(dx)].
\end{eqnarray*}
For every $x\in X$, we have $I_{0,\eps} \subset X_\eps - x$, hence
$\tr \left[ T_0 \qrho (X_\eps - x) \right] > c$. Therefore, whenever $\tr[T\qrho(X)]\neq 0$, we get
\begin{equation*}
\tr\left[ \I_{X_\eps} \left(\I_X (T) \right)\right] > \int_X c \ \tr[T\qrho(dx)] = c\cdot 
\tr\left[ \I_X (T) \right],
\end{equation*}
as claimed.
\end{proof}

\subsection{Discrete version of a position observable}\label{Discrete}

Let $\E$ be an observable and let $\{X_n\}$ be a sequence of disjoint measurable sets such that $\E(\cup_n X_n)=\id$. For each $n$, we denote
\begin{equation}\label{En}
\widetilde{\E}(\{n\})=\E(X_n).
\end{equation}
This equation defines an observable $\widetilde{\E}$. The observable $\widetilde{\E}$ is clearly discrete and we say that $\widetilde{\E}$ is a \emph{discrete version of $\E$}. We emphasize that the properties of $\widetilde{\E}$ depend not only on the observable $\E$ but also on the sequence $\{X_n\}$. Generally, there is no preferential choice of the sequence $\{X_n\}$.

Let $\qrho$ be a position observable. We fix a number $r>0$ and denote $X_n=I_{nr;r}$ for every $n\in\Z$. Then $\qrho(\cup_{n=1}^{\infty} X_n)=\id$, and thus, equation (\ref{En}) defines a discrete version $\widetilde{\qrho}$ of $\qrho$.

\begin{proposition}\label{prop:EQactual}
The effect $\qrho(I_{0;r})$ is actualizable if and only if 
\begin{equation}\label{dia}
{\rm diam}\,\supp \rho < r.
\end{equation}
\end{proposition}

\begin{proof}
Let us first note that $\qrho(I_{0;r})$ has eigenvalue $1$ if and only if there is $X\in \borel$ with positive Lebesgue measure such that $\rho (I_{x,r}) = 1$ $\forall x\in X$. Assume that this condition holds and let $x_1, x_2 \in X$, $x_1\neq x_2$. Then $|x_1 - x_2| < r$, since otherwise $\rho(I_{x_1,r} \cup I_{x_2,r}) = \rho(I_{x_1,r}) + \rho(I_{x_2,r}) = 2$, contradicting $\rho(\R) =1$. We have
\begin{eqnarray*}
1 &=& \rho(I_{x_1,r} \cup I_{x_2,r}) = \rho(I_{x_1,r}) + \rho(I_{x_2,r}) - \rho (I_{x_1,r} \cap I_{x_2,r}) \\
&=& 2 - \rho (I_{x_1,r} \cap I_{x_2,r}),
\end{eqnarray*}
so that $\rho (I_{x_1,r} \cap I_{x_2,r}) = 1$. This implies that $\supp \rho \subset \overline{I_{x_1,r} \cap I_{x_2,r}}$, hence ${\rm diam}\,\supp \rho < r$.

Conversely, suppose that ${\rm diam}\,\supp \rho < r$. Then there exists $\overline{x} \in \R$ and $r^\prime < r$ such that $\supp \rho \subset \overline{I_{\overline{x},r^\prime}}$. If $|x - \overline{x}| < (r-r^\prime)/2$, then $I_{\overline{x},r^\prime} \subset I_{x,r}$, so $\rho (I_{x,r}) = 1$. The claim follows since such $x$'s form a set of positive Lebesgue measure.
\end{proof}

Recalling the discussion after Definition \ref{repeatable}, Propositions \ref{prop:gammaEQ1} and \ref{prop:EQactual} lead to the following result.

\begin{corollary}
The discrete observable $\widetilde{\qrho}$ admits a repeatable instrument if and only if $\gamma(\qrho;1)<r$.
\end{corollary}

We conclude that to obtain a discrete version $\widetilde{\qrho}$ of $\qrho$ which would have a repeatable instrument, one has to choose the partitioning intervals of the outcome space $\R$ strictly bigger than the resolution width $\gamma(\qrho;1)$. A necessary precondition for this is, obviously, that $\gamma(\qrho;1)$ has to be finite.

\section{Joint measurements of position and momentum}\label{Joint}

The problem of joint measurability of position and momentum observables in quantum mechanics has a long history and different viewpoints have been presented. Naturally, an analysis of this problem depends on the definitions of position and momentum observables, and the concept of joint measurability.

In Subsection \ref{Jointdef} we fix the setting of the current discussion. We then show in Subsection \ref{IR} that the product of the resolution widths of jointly measurable position and momentum observables has a positive lower bound. In Subsection \ref{Sequential} we investigate the connection between sequential measurements and joint measurements.

\subsection{Definitions}\label{Jointdef}

The \emph{canonical momentum observable}, denoted by $\p$, is the sharp observable defined as
\begin{equation}\label{fourier}
\p(Y)=\f^{-1}\q(Y)\f,\quad Y\in\borel,
\end{equation}
where $\f$ is the Fourier-Plancherel transformation on $\hi$.
Generally, a momentum observable is defined as a velocity boost
covariant and translation invariant observable. Thus, an observable
$\E:\borel\to\eh$ is a momentum observable if, for all $q,p\in\R$
and $Y\in\borel$, 
\begin{eqnarray}
V_p\E(Y)V_p^* &=& \E(Y+p),\label{covF} \\
U_q\E(Y)U_q^* &=& \E(Y).\label{invF}
\end{eqnarray}
Similarly as in the case of position observables, a probability measure $\nu$ defines a momentum observable $\pnu$ through the formula  
\begin{equation}\label{FY}
\pnu(Y):=\int \nu(Y-y)\ d\p(y),\quad Y\in\borel,
\end{equation}
and all momentum observables have this form.
Since $\q$ and $\p$ satisfy the relation (\ref{fourier}), the results of Section \ref{Position} and \ref{ApproximatePosition} are directly applicable to the case of momentum observables.

A position observable $\qrho$ and a momentum observable $\pnu$ are \emph{jointly measurable} if there exists an observable $\G:\borelr\to\eh$ such that for all $X,Y\in\borel$, 
\begin{equation*}
\qrho(X) = \G(X\times\R),\quad \pnu(Y) = \G(\R\times Y).
\end{equation*}
In this case we say that $\G$ is a \emph{joint observable} of $\qrho$ and $\pnu$, and also that  $\qrho$ and $\pnu$ are the \emph{margins} of $\G$; for motivation and details see, for instance, \cite{Lahti03}.

An observable $\G:\bor{\R^2}\to\lh$ is a \emph{covariant phase space observable} if for all $q,p\in\R$ and $Z\in\bor{\R^2}$,
\begin{equation}\label{G}
U_qV_p\G(Z)V_p^*U_q^*=\G(Z+(q,p)).
\end{equation}
As shown, for instance, in \cite{CaDeTo04}, each covariant phase space observable $\G$ is generated by a unique operator $T\in\sh$ such that $\G=\G_T$,  
\begin{equation}\label{GT}
\G_T(Z)=\frac{1}{2\pi}\int_Z U_qV_pTV_p^*U_q^*\ dqdp,\quad Z\in\borelr.
\end{equation}
We recall that if a position observable $\qrho$ and a momentum observable $\pnu$ have a joint observable, then they also have a joint observable which is a covariant phase space observable; see \cite{CaHeTo05} and \cite{Werner04}.

\subsection{Inaccuracy relation}\label{IR}

By \cite[Corollary 8]{CaHeTo05}, a position observable $\qrho$ and a momentum observable $\pnu$ are jointly measurable if and only if there is a Hilbert space $\ki$ and a vector valued function $\theta\in L^2(\R, dx;\ki)$ such that
\begin{equation}\label{theta}
d\rho(x) = \no{\theta(x)}^2 dx,
\qquad d\nu(y) = \no{\widehat{\theta}(y)}^2 dy,
\end{equation}
where $\widehat{\theta}$ is the Fourier-Plancherel transform of $\theta$.
It is then a consequence of Proposition \ref{prop:EQc0} that for any confidence level $c$, the resolution widths $\gamma(\qrho;c)$ and $\gamma(\pnu;c)$ are strictly positive. The specific form (\ref{theta}) of the probability measures $\rho$ and $\nu$ leads also to the following results, demonstrating the interrelationship between the resolution widths $\gamma(\qrho;c)$ and $\gamma(\pnu;c)$.

\begin{proposition}\label{gamma11}
Let $\qrho$ and $\pnu$ be position and momentum observables which are jointly measurable. Then 
$$
\gamma(\qrho;1)\cdot\gamma(\pnu;1)=\infty.
$$
\end{proposition}

\begin{proof}
Let $\left\{ e_i \right\}$ denote an orthonormal basis of $\ki$. The functions appearing in formula (\ref{theta}) can be written as
\begin{equation}\label{l2decom}
\theta \, = \, \sum_i \theta^i \, e_i\quad {\rm and} \quad 
\widehat{\theta} \, = \, \sum_i \widehat{\theta^i}\, e_i
\end{equation}
where each $\theta^i$ belongs to $\ldl$.

Suppose that $\gamma(\qrho;1)<\infty$. Due to Proposition \ref{prop:gammaEQ1}, we then have $\diam\,\supp\rho<\infty$. This implies that each $\theta^i$ is confined to a bounded interval. Hence, each $\widehat{\theta^i}$ does not vanish on any interval  (see for instance \cite[Section 2.9]{FSI}). Therefore, $\diam\,\supp\nu=\R$. By
Proposition \ref{prop:gammaEQ1} this means that $\gamma(\pnu;1)=\infty$.
\end{proof}

\begin{proposition}\label{ir}
Let $\qrho$ and $\pnu$ be position and momentum observables which are jointly measurable. For any confidence levels $c_1,c_2\in [\frac{1}{2},1]$, we have
\begin{equation}\label{relazione di indeterminazione}
\gamma(\qrho; c_1)\cdot\gamma(\pnu; c_2) \geq 2\pi \left( c_1+c_2 -1\right)^2.
\end{equation}
\end{proposition}

\begin{proof}
Let $\theta\in L^2(\R,dx;\ki)$ be such that (\ref{theta}) holds and let $c_1,c_2\in [\frac{1}{2},1)$. Since the function 
\begin{equation*}
x\mapsto \rho(I_{x;r})=\int_{x-r/2}^{x+r/2} \no{\theta(x')}^2 dx'
\end{equation*}
is continuous and goes to $0$ when $|x|\to\infty$, formula (\ref{gammaEQc}) gives
\begin{equation}\label{kuu}
\gamma(\qrho;c_1)   =   \inf\{r> 0 \mid
\max_{x\in\R}\rho(I_{x;r}) > c_1 \}.
\end{equation}
Similarly,
\begin{equation}\label{luu}
\gamma(\pnu;c_2)  =  \inf\{s > 0 \mid
\max_{y\in\R}\nu(I_{y;s}) > c_2 \}.
\end{equation}

Let $\alpha>\gamma(\qrho;c_1)$ and $\beta>\gamma(\pnu;c_2)$. By formulas (\ref{kuu}) and (\ref{luu}), this means that there exist $\bar{x},\bar{y}\in\R$ such that
\begin{equation}\label{ir1}
\int_{\bar{x}-\alpha/2}^{\bar{x}+\alpha/2} \no{\theta(x)}^2 dx > c_1, \qquad \int_{\bar{y}-\beta/2}^{\bar{y}+\beta/2} \no{\widehat{\theta}(y)}^2 dy > c_2.
\end{equation}
We recall the decomposition (\ref{l2decom}) of $\theta$. As shown in \cite{LaPo61} and \cite{Lenard72}, each $\theta^i\in\ldl$ satisfies
\begin{equation*}
\frac{1}{\no{\theta^i}^2} \left( \int_{\bar{x}-\alpha/2}^{\bar{x}+\alpha/2} |\theta^i(x)|^2 dx+\int_{\bar{y}-\beta/2}^{\bar{y}+\beta/2} |\widehat{\theta^i}(y)|^2 dy \right) \leq 1+\sqrt{\lambda_0},
\end{equation*}
where $\lambda_0$ is the largest eigenvalue of the positive trace class operator $Q(I_{\bar{x};\alpha})P(I_{\bar{y};\beta})Q(I_{\bar{x};\alpha})$. Since $\no{\theta(x)}^2=\sum_i |\theta^i(x)|^2$ and $\sum_i \no{\theta^i}^2=1$, we conclude that
\begin{equation*}
\int_{\bar{x}-\alpha/2}^{\bar{x}+\alpha/2} \no{\theta(x)}^2 dx+ \int_{\bar{y}-\beta/2}^{\bar{y}+\beta/2} \no{\widehat{\theta}(y)}^2 dy\leq 1+\sqrt{\lambda_0},
\end{equation*}
and this with (\ref{ir1}) gives
\begin{equation}\label{ir2}
c_1+c_2<1+\sqrt{\lambda_0}.
\end{equation}
The eigenvalue $\lambda_0$ has the following upper bound:
\begin{equation}\label{ir3}
\lambda_0\leq \tr[Q(I_{\bar{x};\alpha})P(I_{\bar{y};\beta})Q(I_{\bar{x};\alpha})]=\tr[Q(I_{\bar{x};\alpha})P(I_{\bar{y};\beta})]=\frac{\alpha\beta}{2\pi};
\end{equation}
for the last equality, see e.g. \cite{BuLa86}. Thus, combining (\ref{ir2}) and (\ref{ir3}) we get
\begin{equation}
c_1+c_2 <  1 +\sqrt{\frac{\alpha\beta}{2\pi}}.
\end{equation}
Since $\alpha$ and $\beta$ can be chosen arbitrarily close to $\gamma(\qrho;c_1)$ and $\gamma(\pnu;c_2)$, inequality (\ref{relazione di indeterminazione}) follows.

By Proposition \ref{gamma11} we have $\gamma(\qrho;1)\cdot\gamma(\pnu;1)=\infty$. Therefore, to complete the proof it is enough to consider the product $\gamma(\qrho;1)\cdot\gamma(\pnu;c_2)$ for $c_2\neq 1$. We then have 
\begin{equation*}
\gamma(\qrho;1)\cdot\gamma(\pnu;c_2)\geq\gamma(\qrho; c_1)\cdot\gamma(\pnu; c_2) \geq \left( c_1+c_2-1 \right)^2
\end{equation*}
for every $\half\leq c_1<1$. This implies that 
\begin{equation*}
\gamma(\qrho;1)\cdot\gamma(\pnu;c_2) \geq c_2^2,
\end{equation*}
and hence, (\ref{relazione di indeterminazione}) holds.
\end{proof}

If $c_1 = c_2 = \half$, then (\ref{relazione di indeterminazione}) does not give a positive lower bound for the product of the resolution widths. Actually, in this case there exist jointly measurable position observable $\qrho$ and momentum observable $\pnu$ with the product $\gamma(\qrho; \half)\cdot\gamma(\pnu; \half)$ arbitrarily small; this is a consequence of Theorem 2 in~\cite{LaPo61}. Concerning this situation, we note that the related claim in \cite[Proposition 6]{CaHeTo04} is incorrect.

\subsection{Sequential measurements}\label{Sequential}

Let us consider a sequential measurement of a position observable $\qrho$ and a momentum observable $\pnu$. Suppose that $\qrho$ is measured first and the state change is given by a $\qrho$-compatible instrument $\I$, which satisfies the covariance and invariance conditions:
\begin{eqnarray}
U_q\I_X(U_q^*TU_q)U_q^* &=& \I_{X+q}(T), \label{Icov}\\
V_p\I_X(V_p^*TV_p)V_p^* &=& \I_{X}(T), \label{Iinv}
\end{eqnarray}
for every $q,p\in\R,X\in\borel$ and $T\in\sh$. 
We denote by $\I^{\ast}_X:\lh\to\lh$ the dual mapping of $\I_X$. Then $\G$, defined by the condition
\begin{equation}\label{sequentialG}
\G(X\times Y) := \I^{\ast}_X (\pnu(Y)),\quad X,Y\in\borel,
\end{equation}
is the joint observable corresponding to the sequential measurement. (As proved, for instance, in \cite[Theorem 4.5]{Ylinen96}, formula (\ref{sequentialG}) determines a unique observable $\G$ on $\borelr$). 
Also, it follows from (\ref{Icov}) and (\ref{Iinv}) that 
\begin{equation*}
U_q V_p \G(X\times Y) V_p^\ast U_q^\ast = \G(X\times Y + (q,p))
\end{equation*}
for every $q,p\in\R$ and $X,Y\in\borel$, so that $\G$ is a covariant phase space observable. 
Since
\begin{equation*}
\qrho (X) =  \I^{\ast}_X (\id) = \G\left( X\times \R\right),
\end{equation*}
$\rho$ is absolutely continuous with respect to the Lebesgue measure (see the beginning of Subsection \ref{IR}). The other margin
\begin{equation*}
\pnud (Y) :=  \I^{\ast}_\R (\pnu(Y)) = \G\left( \R\times Y \right)
\end{equation*}
depends on the instrument $\I$. Generally, $\pnud$ differs from $\pnu$ since the position measurement disturbs the system.
 
Summarizing, if $\qrho$  admits a covariant and invariant instrument, it is a margin 
of a covariant phase space observable and, in particular, $\rho$ is absolutely continuous with respect to the Lebesgue measure. In the following we show that the converse is also true, namely, if $\rho$ is absolutely continuous then there is a $\qrho$-compatible instrument which is covariant and invariant. Moreover, we show that any covariant phase space observable $\G_T$ generated by a projection $T=\pphi$ can be formed in the previously described manner from a sequential measurement.

Fix a unit vector $\phi\in \hi = \ldl$. For each $\varphi_1,\varphi_2\in \hi$ and $X\in\borel$, let $\I^\phi_X \left(\kb{\varphi_1}{\varphi_2}\right)$ be the integral operator with kernel
\begin{equation*}
K^{\varphi_1,\varphi_2}_X (x,y)  =  \fii_1 (x) \overline{\fii_2 (y)} 
\int \chi_X (z) \overline{\phi(x - z)} \phi (y-z) dz, 
\end{equation*}
that is,
\begin{equation}\label{IphiX}
\left[ \I^\phi_X \left(\kb{\fii_1}{\fii_2}\right) \psi \right] (x) =
\int K^{\varphi_1,\varphi_2}_X (x,y) \psi(y) dy.
\end{equation}
With the notation $\check{f} (x) = f(-x)$ we can write the kernel $K^{\varphi_1,\varphi_2}_X$ in two alternative forms
\begin{eqnarray*} 
K^{\varphi_1,\varphi_2}_X (x,y) & = & \fii_1 (x) \overline{\fii_2 (y)} \left[ (\overline{\phi(\cdot + x)} \check{\chi}_X (\cdot)) \ast \check{\phi} \right] (-y) \\
&=& \fii_1 (x) \overline{\fii_2 (y)} \left[ (\overline{\phi(\cdot + y)} \check{\chi}_X (\cdot)) \ast \overline{\check{\phi}} \right] (-x).
\end{eqnarray*}
Since the convolution of two $L^2$-functions is a bounded function, we conclude that $K^{\varphi_1,\varphi_2}_X$ is in $L^2 (\RR,d^2 x)$ and so, $\I^\phi_X \left(\kb{\fii_1}{\fii_2}\right)$ is a bounded (actually, Hilbert-Schmidt) operator from $\ldl$ into $\ldl$. Moreover, the mapping $\I^\phi_X$ extends by linearity to the space of finite rank operators, which is a dense subspace in $\trh$.

\begin{proposition}\label{str. di fi}
Formula (\ref{IphiX}) determines a unique instrument $\I^\phi$ whose associated observable is $\qrho$, with $d\rho(x) = |\phi(-x)|^2 dx$. The instrument $\I^\phi$ satisfies covariance and invariance conditions (\ref{Icov}) and (\ref{Iinv}).
\end{proposition}

\begin{proof}
We first show that for all $\fii_1,\fii_2 \in \hi$ the operator $\I^\phi_X \left(\kb{\fii_1}{\fii_2}\right)$ is trace class. For all $\psi\in \ldl$, an easy computation gives
\begin{equation}\label{positivita' di I}
\ip{\psi}{\I^\phi_X \left(\kb{\fii_1}{\fii_2}\right)\psi} = \int \chi_X (x) [(\overline{\fii_2} \psi )\ast \check{\phi}] (x) \overline{[(\overline{\fii_1} \psi )\ast \check{\phi}] (x)} dx .
\end{equation}
For each $\fii\in\ldl$, we define the operator $B^{\fii}_{X}$ by the formula
\begin{eqnarray*}
(B^{\fii}_{X} \psi)(x) &=& \chi_X (x) \left[ (\bar{\fii} \psi) \ast \check{\phi} \right] (x) \\
&=& \int B^{\fii}_{X} (x,y) \psi (y) dy,
\end{eqnarray*}
where
$$
B^{\fii}_{X} (x,y) = \chi_X (x) \overline{\fii (y)} \phi (y-x).
$$
Since the kernel $B^{\fii}_{X}(\cdot,\cdot)$ is in $L^2 (\RR,d^2 x)$, the operator $B^{\fii}_{X}$ is Hilbert-Schmidt. By equation (\ref{positivita' di I}) we have
$$
\ip{\psi}{\I^\phi_X \left(\kb{\fii_1}{\fii_2}\right) \psi} = \ip{B^{\fii_1}_{X}\psi}{B^{\fii_2}_{X}\psi} = \ip{\psi}{(B^{\fii_1}_{X})^\ast B^{\fii_2}_{X} \psi}
$$
for all $\psi\in \ldl$, so that $\I^\phi_X \left(\kb{\fii_1}{\fii_2}\right) = (B^{\fii_1}_{X})^\ast B^{\fii_2}_{X}$ is a trace class operator.

Equation (\ref{positivita' di I}) shows that $\I^\phi_X (\pfii)\geq 0$, and hence, using spectral decomposition we conclude that $\I^\phi_X (T) \geq 0$ if $T$ is a positive finite rank operator.

We now show that $\I^\phi_X$ is trace-norm bounded on finite rank operators, so that it uniquely extends to a bounded operator $\I^\phi_X : \trh\frecc\trh$. By decomposition of an operator into its self-adjoint and skew-adjoint parts, we see that it is enough to show that
$\no{\I^\phi_X (T)}_{tr} \leq C \no{T}_{tr}$
for all $T$ self-adjoint and with finite rank (we denote by $\no{\cdot}_{tr}$ the trace class norm). So, let $T$ be finite rank and self-adjoint, and let $T = T_+ - T_-$ be its decomposition into positive and negative parts. Let $T_{\pm} = \sum_{i=1}^{n_{\pm}} \lambda_i^{\pm} P_{\fii_i^{\pm}}$ be the spectral decompositions of the two parts. Since $\I^\phi_X (T_{\pm})$ are positive operators, denoting by $\no{\cdot}_{HS}$ the Hilbert-Schmidt norm, we have
\begin{eqnarray}
\no{\I^\phi_X (T_{\pm})}_{tr} &=&
\tr{\left[\I^\phi_X (T_{\pm})\right]} =
\sum_{i} \lambda_i^{\pm}
\tr{\left[\I^\phi_X (P_{\fii_i^{\pm}})\right]} \nonumber \\
&=& \sum_{i} \lambda_i^{\pm}
\tr{\left[(B^{\fii_i^{\pm}}_{X})^\ast
B^{\fii_i^{\pm}}_{X}\right]}
= \sum_{i} \lambda_i^{\pm}
\no{B^{\fii_i^{\pm}}_{X}}^2_{HS} \nonumber \\
&=& \sum_{i} \lambda_i^{\pm} \int\int \left| B^{\fii_i^{\pm}}_{X} (x,y) \right|^2 dx\,dy  \nonumber \\
&=& \sum_{i} \lambda_i^{\pm} \int \left( \int \chi_X (x) |\phi(y-x)|^2 dx \right) \left| \fii_i^{\pm} (y) \right|^2 dy \nonumber \\
& \equiv & \sum_{i} \lambda_i^{\pm} \ip{\fii_i^{\pm}}{\qrho(X)\fii_i^{\pm}}
= \tr{\left[ T_{\pm} \qrho(X) \right]}, \label{chissa'...}
\end{eqnarray}
where $d\rho(x) = |\phi(-x)|^2 dx$. Hence,
$$
\no{\I^\phi_X (T)}_{tr} \leq \no{\I^\phi_X (T_+)}_{tr} + \no{\I^\phi_X (T_-)}_{tr} \leq \no{\qrho (X)} \left( \no{T_+}_{tr} + \no{T_-}_{tr} \right) = \no{\qrho (X)} \no{T}_{tr},
$$
and the boundedness of $\I^\phi_X$ follows. Note that if $T\in\trh$ is positive, then  equation~(\ref{chissa'...}) implies that
\begin{equation}\label{faccio vedere che e' associato a qrho}
\tr{\left[\I^\phi_X (T)\right]} = \tr{\left[ T \qrho(X) \right]}.
\end{equation}

If $T = \sum_i \lambda_i P_{\fii_i}$ is a positive element of $\trh$ (not necessarily with finite rank) then, by continuity of $\I^\phi_X$ and monotone convergence theorem, equation~(\ref{positivita' di I}) gives
$$
\ip{\psi}{\I^\phi_X (T)\psi} = \int_X \sum_i \lambda_i \left| [(\overline{\fii_i} \psi )\ast \check{\phi}] (x) \right|^2 dx .
$$
Hence, the map $X \mapsto \ip{\psi}{\I^\phi_X (T)\psi}$ is a positive Borel measure and its density with respect to the Lebesgue measure is $\sum_i \lambda_i \left| [(\overline{\fii_i} \psi )\ast \check{\phi}] (x) \right|^2$. In particular, the map $\borel \ni X \mapsto \I^\phi_X (T) \in \lh$ is $\sigma$-additive when $\lh$ is endowed with the weak operator topology. Since $\I^\phi_X (T)$ is positive for each $X$, the map $\borel \ni X \mapsto \I^\phi_X (T) \in \trh$ is $\sigma$-additive in the trace-norm topology. Thus, $\sigma$-additivity in the trace-norm topology for generic $T\in\trh$ then follows.

We have thus shown that $\I^\phi$ is an instrument, whose associated observable is $\qrho$ by equation~(\ref{faccio vedere che e' associato a qrho}).

Finally, from equation (\ref{positivita' di I}) we have
\begin{eqnarray*}
&& \ip{\psi}{U_q V_p \I^\phi_X (V^\ast_p U^\ast_q \pfii U_q V_p )V^\ast_p U^\ast_q \psi} =\\
&& \qquad \qquad = \ip{V^\ast_p U^\ast_q \psi}{\I^\phi_X (V^\ast_p U^\ast_q \pfii U_q V_p )V^\ast_p U^\ast_q \psi} \\
&& \qquad \qquad = \int \chi_X (z) \left|[ \left(\left(\overline{V^\ast_p U^\ast_q \fii} \right) \left( V^\ast_p U^\ast_q \psi \right) \right) \ast \check{\phi}] (z) \right|^2 dz \\
&& \qquad \qquad = \int \chi_X (z) \left|[ \left(\bar{\fii} \psi \right) \ast \check{\phi}] (z+q) \right|^2 dz \\
&& \qquad \qquad = \int \chi_{X+q} (z) \left|[ \left(\bar{\fii} \psi \right) \ast \check{\phi}] (z) \right|^2 dz \\
&& \qquad \qquad = \ip{\psi}{\I^\phi_{X+q} (\pfii) \psi},
\end{eqnarray*}
and so conditions (\ref{Icov}) and (\ref{Iinv}) are satisfied for all $\pfii\in\sh$, hence for all $T\in\sh$.
\end{proof}

Note that if $\phi\in L^\infty (\R) \cap \ldl$, for each $x\in \R$ we can introduce the operator \begin{equation*}
K_x : \ldl \frecc \ldl, \quad [ K_x \psi ] (y) = \overline{\phi (y-x)} \psi (y).
\end{equation*}
We have
$$
[(\bar{\fii} \psi )\ast \check{\phi}] (x) = \ip{K_x \fii}{\psi}
$$
so that equation (\ref{positivita' di I}) can be rewritten as
$$
\ip{\psi}{\I^\phi_X \left(\kb{\fii_1}{\fii_2}\right) \psi} = \int_X \ip{\psi}{K_x \fii_1} \ip{K_x \fii_2}{\psi} dx =
\ip{\psi}{\int_X  K_x \kb{\fii_1}{\fii_2} K_x^\ast \psi} dx .
$$
We thus have for all $T\in \trh$
$$
\I^\phi_X (T) = \int_X  K_x T K_x^\ast dx.
$$
This kind of instrument was introduced in \cite{Davies70} and its properties have been studied in \cite{BuLa90b} and \cite{Ozawa93}. A measurement theoretical model leading to this instrument has been analyzed in \cite{BuLa96b}.

\begin{proposition}
Let $\pnu$ be a momentum observable and let $\G$ be the covariant phase space observable defined via the formula
\begin{equation*}
\G(X\times Y) = \I^{\phi \, \ast}_X (\pnu (Y)), \quad X,Y\in\borel.
\end{equation*}
Then $\G=\G_T$, where $\G_T$ is the observable defined in (\ref{GT}) with
\begin{equation}\label{generating}
T = \int V^\ast_x \pphi V_x \ d\nu(x).
\end{equation}
\end{proposition}

\begin{proof}
We show that $\G=\G_T$ by verifying that $\G(X\times Y) = \G_T(X\times Y)$ for all $X,Y\in\borel$. This is indeed enough since the mapping $(X,Y)\mapsto \G(X\times Y)$ determines a unique observable on $\borelr$; see, for instance, \cite[Theorem 4.5]{Ylinen96}.

For each $\fii\in\hi$, we have (continuing with the notations of the proof of Proposition~\ref{str. di fi})
\begin{eqnarray}
\ip{\fii}{\G(X\times Y) \fii} & = & \tr \left[ \I^\phi_X (\pfii) \pnu(Y) \right] \nonumber \\
& = & \tr \left[ (B^{\fii}_{X})^\ast B^{\fii}_{X} \f^\ast \qnu(Y) \f
\right] = \no{B^{\fii}_{X} \f^\ast \qnu (Y)^\half}^2_{HS}. \label{questa}
\end{eqnarray}
Since the kernel $B^{\fii}_{X}(\cdot,\cdot)$ is in $L^2 (\R^2,d^2 x)$, by Fubini theorem there is a negligible set $Z$ such that $B^{\fii}_{X} (x, \cdot )$ is in $\ldl$ for all $x\in \R \smallsetminus Z$. For such $x$'s and for all $\psi \in \ldl$, we have
\begin{eqnarray*}
\left[ B^{\fii}_{X} \f^\ast \qnu (Y)^\half \psi \right] (x)
& = & \ip{\overline{B^{\fii}_{X} (x, \cdot )}}{\f^\ast \qnu (Y)^\half \psi } \\
& = & \ip{\qnu (Y)^\half \f \overline{B^{\fii}_{X} (x, \cdot )}}{\psi} \\
& = & \ip{\overline{\qnu (Y)^\half  \f^\ast B^{\fii}_{X} (x, \cdot )}}{\psi} \\
& = & \int \nu(Y - y)^\half \left[ \f^\ast B^{\fii}_{X} (x, \cdot ) \right] (y)  \psi (y)\ dy,
\end{eqnarray*}
thus showing that $B^{\fii}_{X} \f^\ast \qnu (Y)^{\half}$ is the integral operator with kernel
\begin{eqnarray*}
\Gamma (x,y) &=& \nu(Y - y)^\half \left[ \f^\ast B^{\fii}_{X} (x, \cdot ) \right] (y) \\
& = & \nu(Y - y)^\half (2\pi )^{-\half} \int e^{iyz} B^{\fii}_{X} (x,z)\ dz \\
& = & (2\pi )^{-\half} \nu(Y - y)^\half \chi_X (x) \int e^{iyz} \overline{\fii (z)} \phi(z-x)\ dz \\
& = & (2\pi )^{-\half} \chi_X (x) \nu(Y - y)^\half \ip{\fii}{V_y U_x \phi}.
\end{eqnarray*}
(here we used the fact that $B^{\fii}_{X} (x,\cdot)$ is in $\lul$ for all $x$ to evaluate explicitly its inverse Fourier transform). So we have
\begin{eqnarray}
&&\no{B^{\fii}_{X} \f^\ast \qnu(Y)^\half}^2_{HS} = \int\int |\Gamma (x,y)|^2 dx \, dy \nonumber \\ 
&&\quad = (2\pi )^{-1}\ip{\fii}{\left( \int\int \chi_X (x) \nu(Y - y) V_y U_x \pphi U_x^\ast V_y^\ast dx \, dy \right) \fii} \nonumber \\
&&\quad = (2\pi )^{-1}\ip{\fii}{\left( \int\int\int \chi_Y (z+y) \chi_X (x) V_y U_x \pphi U_x^\ast V_y^\ast dx \, dy \, d\nu(z) \right) \fii} \nonumber \\
&&\quad = (2\pi )^{-1}\ip{\fii}{\left( \int\int\int \chi_Y (y) \chi_X (x) V_y U_x V_z^\ast\pphi V_z U_x^\ast V_y^\ast dx \, dy \, d\nu(z) \right) \fii} \nonumber \\
&&\quad = \ip{\fii}{\G_T (X\times Y)\fii}. \label{quella}
\end{eqnarray}
Comparing equations (\ref{questa}) and (\ref{quella}), equality $\G(X\times Y) = \G_T(X\times Y)$ follows.
\end{proof}

We recall that the correspondence $T\leftrightarrow\G_T$ between operators in $\sh$ and the covariant phase space observables is one-to-one (see e.g. \cite[Proposition 6]{CaHeTo05}). In particular, an observable $\G_T$ is an extremal point in the convex set of covariant phase observables exactly when $T$ is a projection. This is our motivation for the following statement.

\begin{proposition}
The generating operator $T$ defined in equation (\ref{generating}) is a projection if and only if the momentum observable $\pnu$ is sharp.
\end{proposition}

\begin{proof}
If $\pnu$ is sharp, then $\nu=\delta_{x}$ for some $x\in\R$ and equation (\ref{generating}) gives $T=V_{x}^*\pphi V_{x}=P_{V_{x}^*\phi}$.

Assume then $T$ is a projection, so that it has eigenvalue $1$. Let $\psi$ be a corresponding eigenvector $\psi$ of unit norm. Then
\begin{equation*}
1=\ip{\psi}{T\psi}=\int |\ip{\phi}{V_x\psi} |^2\ d\nu(x),
\end{equation*}
implying that $|\ip{\phi}{V_x\psi} |=1$ for every $x\in\supp\nu$. This shows that if $x,y\in\supp\nu$, then $V_x\psi$ and $V_y\psi$ are proportional to $\phi$, and so $\phi$ is an eigenvector of the operator $V_{x-y}$. But $V_{x-y}$ has eigenvectors only if $V_{x-y}=\id$, i.e., $x=y$. Thus, $\supp\nu$ consists only of one point. 
\end{proof}

Now we turn to the question of the approximate repeatability of the instrument $\I^\phi$. 

\begin{lemma}\label{commute}
Suppose $A\in\lh$ commutes with $\q$. Then $\tr [A \I^\phi_X (T)] = \tr [\I^\phi_X (A T)]$ for all $T\in\sh$.
\end{lemma}

\begin{proof}
It is not restrictive to assume that $A$ is positive, so that there exists a function $\alpha\in L^\infty (\R,dx)$, with $\alpha \geq 0$, such that $(A\psi) (x) = \alpha (x) \psi (x)$; see, for instance, \cite[Section 75]{TLOHS2}. With the notations of the proof of Proposition~\ref{str. di fi}, we have
$$
\tr[A \I^\phi_X (P_\fii)] = \tr \left[ A (B^{\fii}_{X})^\ast B^{\fii}_{X} \right] = \no{A^{\half} (B^{\fii}_{X})^\ast}_{HS}.
$$
The operator $A^{\half} (B^{\fii}_{X})^\ast$ is the integral operator with kernel
$$
K(x,y) = \alpha(x)^{\half} \chi_X (y) \fii (x) \overline{\phi(x-y)},
$$
so that
\begin{eqnarray*}
\no{A^\half (B^{\fii}_{X})^\ast}_{HS} &=& \int\int | K(x,y) |^2 dx dy \\
&=& \int \alpha(x) \left(\int \chi_X (y) | \phi(x-y) |^2 dy \right) | \fii (x) |^2 dx \\
&=& \int \alpha(x) \rho(X -x) | \fii (x) |^2 dx = \ip{\fii}{\qrho(X) A \fii} \\
&=& \tr [\I^\phi_X (A P_\fii)],
\end{eqnarray*}
where $d\rho (x) = |\phi(-x)|^2 dx$. This proves the lemma for $T = P_\fii$. The claim for general $T\in\sh$ then follows. 
\end{proof}

\begin{proposition}\label{repphi}
The instrument $\I^\phi$ has the following properties.
\begin{itemize}
\item[(i)] $\I^\phi$ is $\eps$-repeatable for any $\eps > 2\cdot \gamma(\qrho;1)$.
\item[(ii)] If $\gamma(\qrho;1)=\infty$, there is no $\eps >0$ and $\half \leq c \leq 1$ such that $\I^\phi$ is $(\eps,c)$-repeatable.
\end{itemize}
\end{proposition}

\begin{proof}
\begin{itemize}
\item[(i)] Let $\eps > 2\cdot \gamma(\qrho;1)$, which by Proposition \ref{prop:gammaEQ1} means that $\eps > 2\cdot \diam\,\supp \rho$. Using Lemma \ref{commute} we get
\begin{eqnarray}
\tr [\I^\phi_{X_\eps} (\I^\phi_X (P_\fii))] &=&
\tr [\qrho(X_\eps) (\I^\phi_X (P_\fii))] = 
\tr [\I^\phi_X (\qrho(X_\eps) P_\fii)] \nonumber\\
&=& \tr [\qrho(X)\qrho(X_\eps) P_\fii] \nonumber\\
&=& \int \rho(X_\eps -x) \rho(X -x) |\fii(x)|^2 dx. \label{equazione prima}
\end{eqnarray}
On the other hand
\begin{eqnarray}
\tr [\I^\phi_X (P_\fii)] &=& \tr [\qrho(X) P_\fii] \nonumber\\ 
&=& \int \rho(X -x) |\fii(x)|^2 dx. \label{equazione seconda}
\end{eqnarray}
If $\rho(X -x) > 0$ for some $x\in\R$, then $(X -x) \cap \supp \rho \neq \emptyset$, so that $\supp \rho\subset X_\eps -x $, and then $\rho(X_\eps -x) = 1$. The claim then follows comparing equations (\ref{equazione prima}) and (\ref{equazione seconda}).

\item[(ii)] Assume that $\I^\phi$ is $(\eps,c)$-repeatable. As noticed in (i), Lemma \ref{commute} implies that 
\begin{equation*}
\tr [\I^\phi_{X_\eps} (\I^\phi_X (T))] = \tr [\qrho(X_\eps)\qrho(X) T]
\end{equation*}
for any $T\in\sh$.
Hence, the requirement that
$$
\tr [\I^\phi_{X_\eps} (\I^\phi_X (T))] > c\cdot\tr [\I^\phi_X (T)]\quad \forall T\in\sh
$$
is equivalent with 
$$
\ip{\psi}{\qrho(X_\eps)\qrho(X)\psi} > c\ip{\psi}{\qrho(X)\psi}\quad \forall \psi\in\hi,\psi\neq 0.
$$
This means that $\rho(X_\eps -x) \rho(X -x) > c\rho(X -x)$ for almost all $x$. So, we must have $\rho(X_\eps -x) > c$ for almost all $x$ such that $\rho(X -x) > 0$, and, since $x\mapsto \rho(X-x)$ is a continuous function, this amounts to $\rho(X_\eps -x) > c$ for all $x\in A := \{ x \mid \rho(X -x) > 0 \}$. If $\gamma(\qrho;1)=\infty$, then $\diam\,\supp \rho = \infty$. Take $X = I_{0,r}$ with $r>0$, in which case the set $A$ is unbounded.  Since $\rho(X_\eps -x) > c$ for all $x\in A$, this is in contradiction with $\rho(\R)=1$.
\end{itemize}
\end{proof}

Finally, we note that Proposition \ref{gamma11} together with Proposition \ref{repphi} lead to the following trade-off relation between the approximate repeatability of a position measurement and the corresponding momentum disturbance. Consider again the sequential measurement procedure described in the beginning of this subsection, where a measurement of $\qrho$ is followed by a measurement of $\pnu$. If the $\qrho$-compatible instrument $\I^\phi$ is  $(\eps,c)$-repeatable for some $\eps$ and $c$, then the position measurement disturbs the system in such a way that the actually measured momentum observable $\pnud$ has $\gamma(\pnud;1)=\infty$.

\section*{Acknowledgements}

This work has mainly been done during a visit of T.H. at Dipartimento di Fisica, Universit\`a di Genova. The hospitality of Dipartimento di Fisica and the financial support by Helsingin Sanomain 100-vuotiss\"a\"ati\"o are gratefully acknowledged. Authors want to thank Paul Busch and Pekka Lahti for helpful comments on an earlier version of this paper and for their proposed improvement in Proposition \ref{ir}.

\end{document}